\documentclass[a4paper,11pt]{article}
\pdfoutput=1 % if your are submitting a pdflatex (i.e. if you have
\usepackage{jcappub} % for details on the use of the package, please
\usepackage[T1]{fontenc} % if needed

\usepackage{booktabs} % for the use of \midrule

\title{\boldmath Detecting light dark matter with prompt-delayed events in neutrino experiments}

\author[a]{Yuanlin Gong,}
\author[a]{Feiran Lin,}
\author[a,b]{Ning Liu,}
\author[a]{Liangliang Su,}
\author[a,b]{and Lei Wu,}

\affiliation[a]{Department of Physics and Institute of Theoretical Physics, Nanjing Normal University,\\Nanjing, 210023, China}
\affiliation[b]{Nanjing Key Laboratory of Particle Physics and Astrophysics,\\ Nanjing, 210023, China}
\emailAdd{yuanlingong@nnu.edu.cn}
\emailAdd{feiranlin@njnu.edu.cn}
\emailAdd{liuning@njnu.edu.cn}
\emailAdd{liangliangsu@njnu.edu.cn}
\emailAdd{leiwu@njnu.edu.cn}

\abstract{ We demonstrate that the prompt-delayed signals induced by knockout neutrons from quasi-elastic scattering in liquid scintillator neutrino experiments provide a new avenue for detecting light dark matter. As an illustration, we consider the detection of atmospheric dark matter, showing that the constraint on the DM–nucleon interaction from KamLAND is approximately one order of magnitude more stringent than those derived from elastic nuclear recoil signals in dark matter direct detection experiments. Furthermore, larger-volume neutrino detectors such as JUNO are expected to further enhance the sensitivity to light dark matter through quasi-elastic scattering. }

\begin{document}
\maketitle
\flushbottom

\section{Introduction}
\label{sec:intro}

Numerous compelling gravitational evidences from astrophysical and cosmological observations confirms the existence of dark matter (DM) in the universe. Despite extensive efforts over several decades, no conclusive non-gravitational interaction signals have been observed in numerous DM detection experiments for the most popular DM candidates--weakly interacting massive particles (WIMPs) ~\cite{Roszkowski:2017nbc,CDEX:2020tkb,PandaX-4T:2021bab,XENON:2023cxc}. In addition to advancing DM detection techniques, increasing attentions have been directed toward non conventional WIMP candidates, such as light DM~\cite{Essig:2011nj,Essig:2015cda,Kouvaris:2016afs,Schutz:2016tid,Hochberg:2016sqx,Knapen:2017xzo,Ibe:2017yqa,Dolan:2017xbu,Hochberg:2019cyy,Flambaum:2020xxo,Kahn:2021ttr,Wang:2021oha,Bell:2021ihi,Mitridate:2022tnv,Gu:2022vgb,Li:2022acp,Bhattiprolu:2023akk,Essig:2024wtj,Bhattiprolu:2024dmh,Balan:2024cmq,Balan:2025uke}.

The searches for the light DM in the direct detection are usually hampered by the small recoil energy. Therefore, the relativistic light DM that produced from the interactions of DM with the high energy astrophysical objects has attracted great attentions, including cosmic ray up-scattering DM~\cite{Bringmann:2018cvk,Ema:2018bih,Guo:2020oum,Wang:2019jtk,Ge:2020yuf,Ema:2020ulo,Xia:2020apm,Elor:2021swj,Bell:2021xff,Feng:2021hyz,PandaX-II:2021kai,Maity:2022exk,Darme:2022bew,Liang:2024xcx,Ghosh:2024dqw}, supernova neutrino-boosted DM~\cite{Das:2021lcr,Jho:2021rmn,Lin:2022dbl,Lin:2024vzy,Sun:2025gyj,DeRomeri:2023ytt}, atmospheric DM~\cite{Alvey:2019zaa,Su:2020zny,Arguelles:2022fqq,Du:2022hms}. So far, the primary observable signals in DM direct detection experiments have originated from the elastic nuclear recoil events between DM and target nuclei. However, recent studies, as indicated in Refs~\cite{Alvey:2022pad, Su:2022wpj, PandaX:2023tfq, Su:2023zgr}, have shown that relativistic DM-nucleus scattering is dominated by inelastic scattering rather than elastic scattering. Moreover, the inelastic scattering between DM and target nuclei can generate additional observable signals like the de-excitation spectrum of the excited nucleus, which typically occur on the MeV scale~\cite{Dutta:2024kuj}. Significantly, while they are beyond the primary region of interest for traditional DM direct detection experiments, they fall well within the detection capabilities of neutrino experiments. In fact, the inelastic scattering processes, especially the quasi-elastic scattering (QES), are important signatures in large water-Cherenkov and liquid-scintillator (LS) detectors. They usually produce excited daughter nuclei, nucleons, and $\gamma$-rays, which have been utilized in studies such as the diffuse supernova neutrino background~\cite{KamLAND:2011bnd,Mollenberg:2014pwa,Wei:2016vjd,JUNO:2022lpc}, invisible decay modes of neutron~\cite{JUNO:2024pur}, the de-excitation and neutron-capture $\gamma$ rays (``$\gamma$ + n'' pair) from QES process for detecting light DM~\cite{Choi:2024ism} and DM annihilation to neutrinos~\cite{Chauhan:2021fzu,JUNO:2023vyz}. 

\begin{figure}
\centering
\includegraphics[scale=0.3]{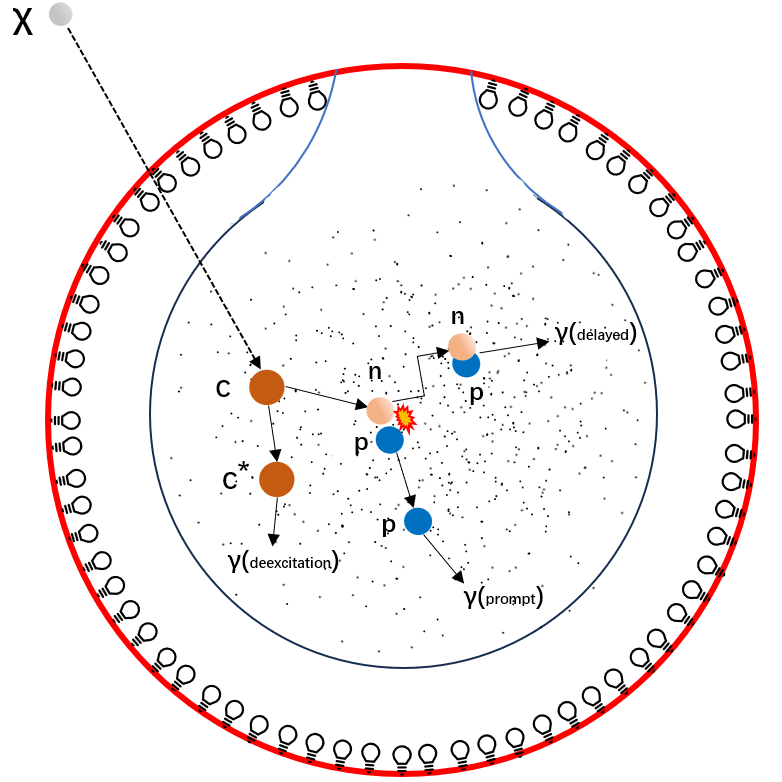}
\caption{A sketch of the quasi-elastic scattering of a relativistic DM with a carbon nucleus in the liquid scintillator detector of neutrino experiment, $\chi + A \rightarrow \chi +(A-1)^{\star} + n$. The knockout neutron will lead to the prompt signal through the elastic scattering process, $n + p \to n + p$, where the recoiling proton emits scintillation light as it passes through medium. The delayed signal is caused by the radiative capture of knockout neutron, $n+p\to d + \gamma$, emitting 2.2 $\mathrm{MeV}\ \gamma$-ray. Besides, the excited residual nucleus can also produce an observable signal through its de-excitation.}
\label{sketch}
\end{figure}

In this work, we propose a novel approach to detect light dark matter in liquid-scintillator neutrino experiments by exploiting the prompt-delayed signals induced by knockout nucleons from the DM-nucleus quasi-elastic scattering process $\chi + A \rightarrow \chi + (A-1)^{\star} + n$, as illustrated in Fig.~\ref{sketch}. LS detectors like KamLAND has excellent energy resolution and neutron tagging capability, enabling the precise measurements of such processes at the MeV scale~\cite{KamLAND:2011bnd,KamLAND:2021gvi,Cheng:2024uyj}. The correlation of prompt and delayed signals can significantly suppress accidental backgrounds and thus serve as a robust signature for probing neutrino properties \cite{KamLAND:2011bnd,KamLAND:2021gvi,Borexino:2013zhu} and new physics. We take atmospheric DM as a benchmark model of relativistic DM and derive the upper limit on the DM-nucleon scattering cross section using the quasi-elastic scattering data from the KamLAND experiment. We find that our bound can be several tens of times more stringent than those from elastic scattering in direct detection experiments. Furthermore, these limits can be improved by about one order of magnitude in upcoming larger-volume neutrino experiments like JUNO. This reveals the great potential of the QES process in neutrino experiments for probing relativistic light DM.

\section{Prompt-delayed signal from QES of atmospheric DM}

Relativistic light dark matter can arise from the inelastic scattering of cosmic rays, dominated by protons and helium, with Earth’s nitrogen-rich atmosphere~\cite{Alvey:2019zaa}, under the assumption of DM interactions with the Standard Model (SM). This assumption, fundamental to direct detection experiments, necessitates the existence of such a DM component. In principle, the secondary cosmic-ray can also accelerate the light DM. In our study, we neglect these contributions, resulting in a smaller flux and a more conservative result. To illustrate concretely, we consider the hadrophilic dark sector, which introduces a singlet scalar mediator $S$ and a Dirac fermion DM $\chi$~\cite{Batell:2018fqo}. The relevant interactions are given by,
\begin{equation}
\begin{split}
\mathcal{L}_S \supset g_\chi S\bar{\chi}_L\chi_R+g_uS\bar{u}_Lu_R + \text{h.c.},\label{eq2}
\end{split}
\end{equation}
where the couplings of mediator $S$ with the DM and up-quark are denoted by $g_\chi$ and $g_u$, respectively. Thus, the coupling of mediator $S$ with the neutron and proton can be written as $g_{nS} = 0.012g_u m_n/m_u$ and $g_{pS} = 0.014g_u m_p/m_u$~\cite{Durr:2015dna}, where $m_n$, $m_p$ and $m_u$ are the masses of neutron, proton and $u$ quark, respectively. Due to Eq.~\ref{eq2}, a huge amount of light mesons $X$ can be produced from the inelastic scattering of high-energy protons with nitrogen, $p+N \to X$, which can decay into light DM particles via the on-shell $S$ mediator, $X \to \pi + S(\to\chi\bar{\chi})$. As the mediator decay on-shell, the mediator mass satisfies $2m_\chi<m_S<m_X - m_\pi$. The decays of these light mesons are constrained by the beam-dump experiments, such as the E787/949~\cite{E787:2002qfb,E787:2004ovg,E949:2007xyy} and MiniBooNE ~\cite{MiniBooNEDM:2018cxm}, and also impose a strong constrain on $g_u$ and $m_S$. {Especially, Big Bang Nucleosynthesis has imposed the constraint $m_S > 20$ MeV \cite{Batell:2018fqo}}. Given the existing experimental constraints, we focus on the decay process $\eta \to \pi \chi \bar{\chi}$ as its branching ratio is still allowed to be relatively large, BR$(\eta \to \pi +invisible)\ \lesssim \ 1\times 10^{-4}$~\cite{ParticleDataGroup:2024cfk}. {Thus, we take $m_S =50,\,100,\, 300$ MeV and BR$(\eta \to \pi \chi \bar{\chi})\ \simeq \ 6\times 10^{-7},\, 6\times 10^{-7},\, 1\times 10^{-5}$ respectively, as our benchmarks. The smaller branching ratios for the lighter mediators arise from the stronger experimental constraints from E787/949~\cite{E787:2002qfb,E787:2004ovg,E949:2007xyy}}. And the branching ratio for the decay of $S$  into DM is taken to be 1.
Then, the differential flux of atmospheric DM is insensitive to the mediator mass and can be written as
\begin{equation}
\begin{aligned}
\frac{\text{d}\Phi_{\chi}}{\text{d}E_{\chi}}=&D\int \text{d}T_p \sigma_{pN\to \eta} \frac{\text{d}\Phi_p(h_{max}, T_p)}{\text{d}T_p} \text{BR}(\eta \to  \pi^0\chi\bar{\chi}) \frac{\text{d}\Gamma_{\eta \to  \pi^0\chi\bar{\chi}}}{{\Gamma_{\eta \to  \pi^0\chi\bar{\chi}}} \text{d}E_\chi},
\label{eq:flux}
\end{aligned}
\end{equation}
where $\sigma_{pN\to \eta} \text{d}\Phi_p/\text{d}T_p = \sigma_{pN}\text{BR}(pN\to \eta)\text{d}\Phi_p/\text{d}T_p$ describes the production of $\eta$. $\Phi_p(h_{max},T_p)$ denotes the height-independent flux of high-energy protons, given at maximum height $h_{max}$ from ground level. We take the inelastic proton-nitrogen cross-section $\sigma_{pN} \approx 255$ mb as constant in the relevant energy range of this work and simulate the production of $\eta$ from the scattering of protons with nitrogen by CRMC package \cite{Pierog:2013ria,ulrich_2021_5270381}. $T_p$ and $E_\chi$ represent the kinetic energy of the incident protons and the energy of the produced DM, respectively. For the on-shell $S$ mediator, the normalized differential decay rate ${\text{d}\Gamma_{\eta \to  \pi^0\chi\bar{\chi}}}/({\Gamma_{\eta \to  \pi^0\chi\bar{\chi}}}{\text{d}E_\chi}) $ can be expressed as the product of two sequential two-body decays (see Ref.~\cite{Arguelles:2019ziu,Xotta:2023jzg}). The total dilution factor $D$ is given by
\begin{equation}
\begin{split}
	D=&\int^{h_{max}}_0 \text{d}h(R_E+h)^2\int^{2\pi}_0 \text{d}\phi\int^{+1}_{-1}\text{d}\cos\theta \frac{y_p(h)}{s^2_d(h,\theta)}n_N(h),
\end{split}
\end{equation}
where $R_E$ is the Earth radius, $s_d$ is the line of sight distance between the point of DM production and the detector and $n_N(h)$ denotes the number density of nitrogen at height $h$ over the ground level. $y_p (h)= \text{exp}(-\sigma_{pN}\int^{h_{max}}_h\text{d}\tilde{h}n_N(\tilde{h}))$ is the dilution factor of the cosmic rays in the atmosphere, with $h_{max} = 180$ km. Note that the atmospheric DM flux may be attenuated by interactions with the Earth before reaching the detector. However, for the range of interaction strengths considered in this work, such an attenuation effect is negligible~\cite{Su:2022wpj}. Moreover, the kinetic energy of atmospheric DM near the detector is concentrated at hundreds of MeV~\cite{Su:2020zny}. In this energy range, quasi-elastic scattering with nuclei dominates over elastic scattering~\cite{Su:2022wpj}, making it readily detectable by large-volume neutrino experiments.

In our study, we consider the KamLAND experiment. The QES of the atmospheric DM with the carbon nuclei occurs via the process,
\begin{equation}
\chi + ^{12}\text{C} \to \chi + n +^{11}\text{C}^{*}.\label{eq:C_QES}
\end{equation}
The knockout neutron scattering with the surrounding medium leads to the prompt energy deposition signal ( $\sim$ few ns), where the $n + p \to n + p$ process is dominated due to the nearly equal masses. The recoiling proton, which inherits most of the neutron’s energy, subsequently deposits energy by producing scintillation light in the LS detector. {Meanwhile, the slowed-down neutrons are further moderated to sub-thermal energies by multiple elastic scatterings with protons and eventually captured by $^1\mathrm{H}$ via the process $n+p\to d + \gamma$, emitting a 2.2~MeV $\gamma$ ray, which constitutes the so-called delayed signal. The capture probability is as high as 99.5\% \cite{KamLAND:2009zwo} and the capture time is 210 $\mu$s on average. We therefore assume one-to-one correspondence between prompt events and delayed neutron-capture events. In the experimental data analysis and events selection of QES, we use the correlation of prompt and delayed signals in time and space, typically within 1000~$\mu$s (mean $\sim210$~$\mu$s) and 160~cm (mean $\sim$60~cm), to efficiently reduce the backgrounds as demonstrated in~\cite{KamLAND:2021gvi}.} This prompt-delayed signal is beneficial for the QES detection of relativistic light DM in LS detector. It is worth noting that $\chi$–H scattering can also produce fast protons that may subsequently eject neutrons from carbon nuclei, e.g., via $^{12}$C(p,n)$^{12}$N, leading to the same experimental signature. However, this two-step channel is expected to be subdominant compared to the direct $\chi$–C interaction because of 
% the smaller number densities in LS \cite{KamLAND:2011bnd,JUNO:2015zny,Meighen-Berger:2023xpr}, 
the absence of $A^2$ coherent enhancement in $\chi$--H scattering and the additional suppression from the required secondary hadronic interaction. In addition to the process of ``neutron-only'' knockouts in Eq.~\ref{eq:C_QES}, the ``proton-only'' knockouts can also occur in the QES process. However, LS detectors are difficult to distinguish it from the elastic scattering signal due to the absence of delayed signals, leading to an irreducible backgrounds. Therefore, our analysis will focus exclusively on the ``neutron-only'' knockout signals. Moreover, the combined signal from the de-excitation photon and the delayed capture of the knockout neutron can also offer a promising channel for detecting light DM, as shown in~\cite{Choi:2024ism}.

In the calculation of QES cross section, we assume the target nucleus as a collection of individual nucleons and the independent evolution of the particles produced at the interaction vertex and the recoiling $(A-1)$-nucleon system~\cite{Benhar:2005dj,benhar2008inclusive}. This impulse approximation performs well for the high momentum transfer ($|\vec{q}|>350\ \mathrm{MeV}$). Then, we can obtain the differential cross section of Eq.~\ref{eq:C_QES}, 
\begin{equation}
\begin{aligned}
	\frac{\text{d}^2\bar{\sigma}_{\chi C}}{\text{d} E_{\vec{p^\prime}} \text{d}\Omega}
	 =&  \frac{(A-Z)\bar{\sigma}_n m_S^4}{16 \pi \mu_n^2(Q^2+m_S^2)^2}\int \text{d}^3\vec{p} \text{d}E\frac{m_{n}^2}{ E_{\vec{p}}E_{\vec{p}^\prime}} \frac{|\vec{k}^{\prime}|}{|\vec{k}|} \\
	&\times P_n(\vec{p}, E)\delta(\omega -E +m_n -E_{\vec{p}^\prime}){\mathcal{X}}^{S} \tilde{W}_n^{S} ,\label{eq:QES}
\end{aligned}
\end{equation}
where $k = (E_{\chi}, \vec{k})$ and $k^{\prime} = (E_{\chi}^{\prime}, \vec{k}^{\prime})$ are the four-momentum of atmospheric DM before and after scattering, respectively, in the rest frame of carbon nuclei, as well as $Q^2 = - (k-k^{\prime})^2$. The four-momentum of initial and knockout neutron at the interaction vertex are denoted by $p = (E_{\vec{p}},\vec{p})$ and $p^{\prime} = (E_{{\vec p}^{\prime}},{\vec p}^{\prime})$. The spectral function $P_n(\vec{p}, E)$ characterizes the distribution of neutron in the plane defined by their momentum $|\vec{p}|$ and removal energy $E$, which can be modeled using the local density approximation~\cite{Benhar:1994hw}. Furthermore, we define a spin-independent DM-nucleon seattering cross section $\bar{\sigma}_n \equiv g_\chi^2g_{nS}^2\mu_n^2/\pi m_S^4$, where $\mu_n$ is the reduced mass between DM particle and nucleon.  The DM tensor ${\mathcal{X}}_{S}$ and hadronic tensor $\tilde{W}_n^{S}$ are written as,
\begin{equation}
\begin{aligned}
	\mathcal{X}^{S} = &\bar{\sum}\langle\chi |j_\chi^S| \chi^\prime \rangle \langle\chi^\prime |j_\chi^S| \chi \rangle = 4m_\chi^2 +Q^2; \\
	\tilde{W}_n^{S} = &\bar{\sum}\langle N,\vec{p} |j_n^S| x,\vec{p}+\vec{q} \rangle \langle\vec{p}+\vec{q},x |j_n^S| N,\vec{p} \rangle \\
= & (1 - \frac{\tilde{q}^2}{{4m_n^2}})F_S^2(Q^2),
\end{aligned}
\end{equation}
where $\tilde{q} \equiv (\tilde{\omega},\vec{q})$ is the modified four-momentum transfer to neutrons, and $\tilde{\omega}=E_{\vec{p}^\prime}-E_{\vec{p}} = \omega - E + m_n - E_{\vec{p}}$. The scalar nucleon form factor $F_S(Q^2)$ used in this work is taken from Ref.~\cite{Schardmuller:2013fqa}. In addition, the Pauli blocking and dynamical final-state interactions (FSI) of the outgoing particles can also impact the emission of low-momentum neutrons~\cite{Bodek:2018lmc, MINERvA:2019ope, Prasad:2024gnv}. We account for Pauli blocking by introducing the step function $\theta(|\vec{p} + \vec{q}| - \bar{p}_F)$ in Eq.~\ref{eq:QES}, where $\bar{p}_F = 221~\text{MeV}$ denotes the average Fermi momentum of carbon nuclei. The FSI effects can be included through nuclear potentials, such as the nuclear optical potential $U_{\text{opt}}(\vec{p} + \vec{q})$ and the Coulomb potential. In the case of the neutral neutron considered in Eq.~\ref{eq:QES}, we account for the influence of the nuclear optical potential, which can be parameterized as~\cite{Bodek:2018lmc, MINERvA:2019ope},
\begin{equation}
U_{\text{opt}} = \text{min}[0,-29.1+(40.9/\text{GeV}^2)(\vec{p}+\vec{q})^2]\ \text{MeV}.
\end{equation}
This will modify the kinematics of the struck particle. Consequently, the final knockout neutron energy is given by $E_{\vec{p}^\prime}^f =E_{\vec{p}^\prime} - |U_{\text{opt}}|$. More sophisticated implementation of the FSI can be simulated by the neutrino Monte Carlo generator like GENIE and NuWro as shown in \cite{Bodek:2018lmc, MINERvA:2019ope,Meighen-Berger:2023xpr}, which can take the knockout of more additional nucleons and channels such as charge exchange multi-nucleon knockout into account. Similarly, FSI with the $\chi$ interacting
with a proton could result in the emission of a neutron. These effects can modestly alter the relevant cross-section \cite{MINERvA:2019ope,Cheng:2020oko,Prasad:2024gnv} and are referred to further extension of this work.

With Eqs.~\ref{eq:flux} and \ref{eq:QES}, the expected differential events numbers can be calculated as
\begin{equation}
	\frac{\text{d} \mathcal{N}}{\text{d} E_{\vec{p}^\prime}^f} = n_C H\int_{E_{\chi}^{min}} ^{E_{\chi}^{max}} \epsilon \text{d}E_\chi \frac{\text{d}\Phi_\chi}{\text{d} E_\chi} \frac{\text{d}\bar{\sigma}_{\chi C}}{\text{d}E_{\vec{p}^\prime}^f} ,
\label{eq:diff_event}
\end{equation}
where $H$ is the experimental exposure. $n_C = 4.29 \times \ 10^{31} \;\text{kt}^{-1}$ denote the number density of carbon nuclei in KamLAND, and the detection efficiency (livetime fraction $\times$ analysis efficiency) is assumed to be $\epsilon= 0.58$~\cite{Meighen-Berger:2023xpr}.
\begin{figure}
	\centering
	\includegraphics[width=0.465\linewidth]{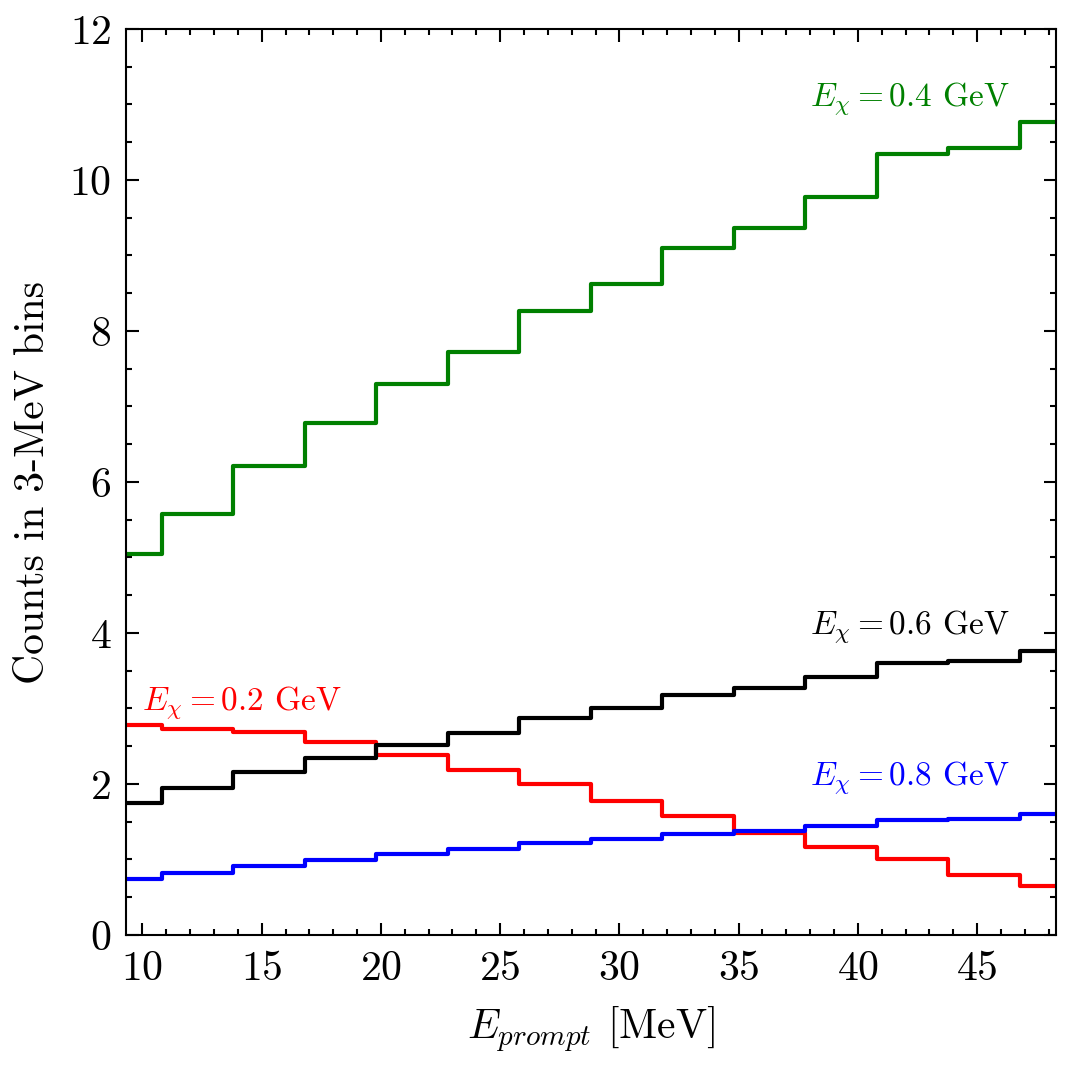}	
	\includegraphics[width=0.5\linewidth]{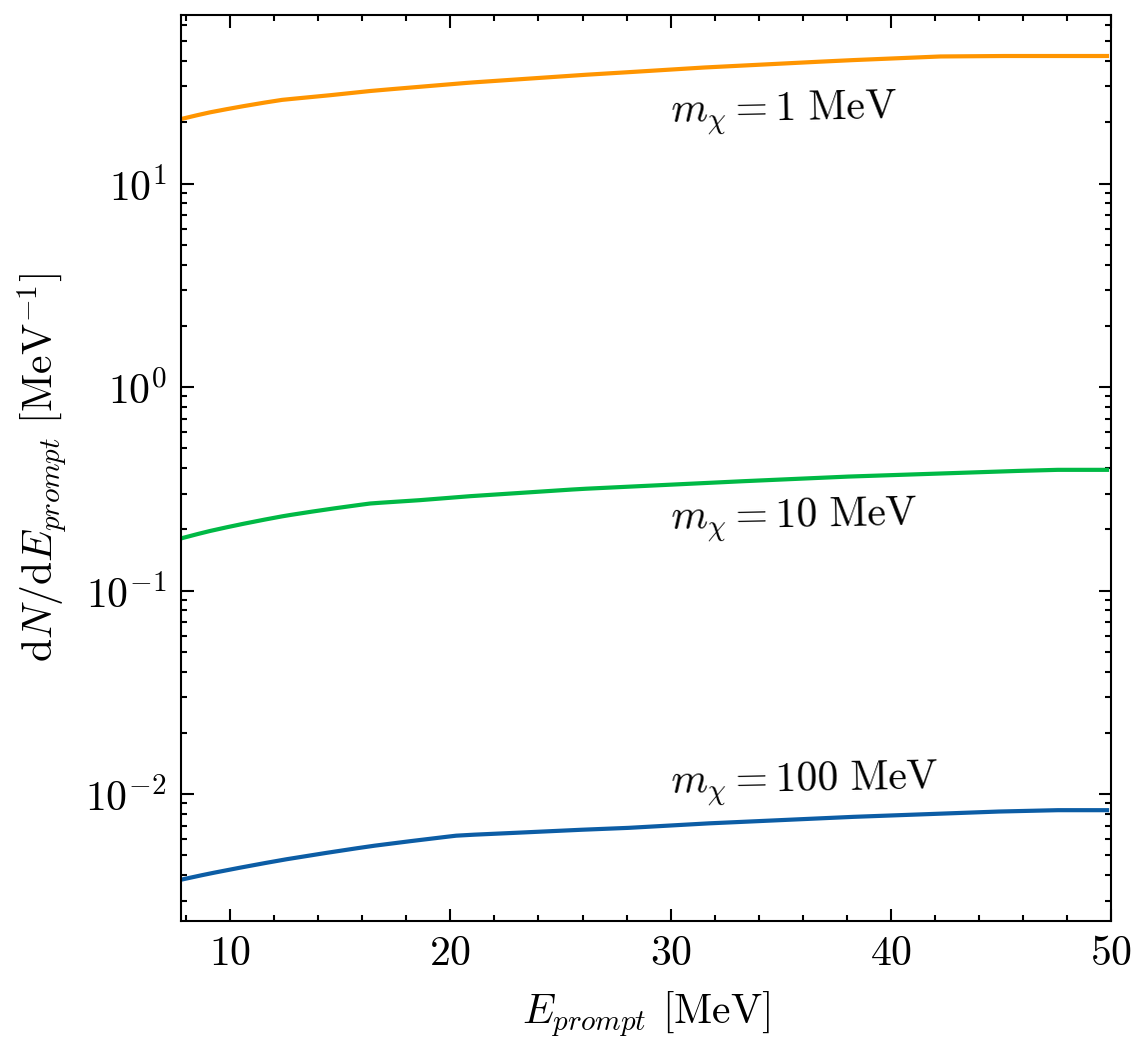}
	\caption{\textbf{Left:} The expected counts in 3--MeV bins for $m_\chi = 0.1$\ MeV DM with energy $E_\chi = 0.2, 0.4, 0.6, 0.8$\ GeV. \textbf{Right}: The expected differential deposited  energy spectrum of the prompt signals for the QES of DM-nucleus at $m_\chi = 1, 10, 100$\ MeV. In both plots, we take $\bar{\sigma}_n \simeq 1\times 10^{-33}$ cm$^{2}$ and $m_S = 0.3$\ GeV. } 
	\label{Kam}
\end{figure}
In the left plane of Fig. 2, we show the predicted QES event induced by 0.1~MeV atmospheric DM in the KamLAND experiment, binned by deposited energy, for various energies $E_\chi = 0.2, 0.4, 0.6, 0.8$\ GeV. {We find that, in the low deposited energy region of interest in this work, the dominant contribution comes from atmospheric DM with incident energies between $200$ MeV and $1$ GeV. Therefore, we take $E^{min}_\chi = 200$ MeV and $E^{max}_\chi= 1$ GeV in Eq. (2.8).} This is because higher-energy atmospheric DM is suppressed by kinematic constraints and spectral functions. For DM with kinetic energy $E_\chi = 0.2$\ GeV, the differential rate at large prompt deposited energies is suppressed by the Pauli blocking effect, $\theta(|\vec{p} + \vec{q}| - \bar{p}_F)$. At even lower DM energies, this suppression becomes more pronounced, as the reduced kinetic energy further limits the available momentum transfer, making it increasingly difficult to efficiently trigger the QES process. As mentioned before, regarding the energy reconstruction of the experiments, the observable deposited energy is from the energetic proton produced via the dominant elastic scattering $n + p \to n + p$ with the knockout neutron. Due to the nonrelativistic nature of the energetic proton, the deposited energy of the scintillation light is lower than the proton kinetic energy $T^{recoil}_p \approx E_{\vec{p}^\prime}^f - m_n$, unlike the case involving the relativistic electron with same kinetic energy. This phenomenon, known as the quenching effect, can be described by Birk’s Law~\cite{birks1951scintillations,Chou:1952jqv},
\begin{equation}
E_{prompt}  = \int_{0}^{T^{recoil}_p}\frac{\text{d}T}{1+k_B\langle \text{d}E/\text{d}x\rangle +k_C\langle \text{d}E/\text{d}x\rangle^2}, \label{quenching}
\end{equation}
where $E_{prompt}$ is the deposited energy of prompt signal. The function $\langle \text{d}E/\text{d}x\rangle$ denotes the average energy loss of a proton in the detector material, which depends on the detector composition. For the KamLAND (JUNO) experiment, the energy loss rate is given by $\langle \text{d}E/\text{d}x\rangle$ $\equiv$ $0.85(0.88)\langle \text{d}E/\text{d}x\rangle_C$+$0.15 (0.12) \langle \text{d} E/\text{d} x \rangle_H$, where $\langle \text{d} E / \text{d} x \rangle_{H, C}$ are energy loss rate on  hydrogen and carbon, respectively, taken from the PSTAR program~\cite{PSTAR}. Meanwhile, the parameter $k_B = 7.79(6.5)  \times  10^{-3} ~\mathrm{g/ cm^2/ MeV}$ and $k_C = 1.64(1.5) \times 10^{-5} ~\mathrm{(g/ cm^2/MeV)^2}$ are the Birks' constants for KamLAND (JUNO) experiment~\cite{Yoshida:2010zzd,Meighen-Berger:2023xpr}.

Therefore, with Eqs.~\ref{eq:diff_event} and~\ref{quenching}, we obtain the differential number of QES events as a function of the (prompt) deposited energy ${\text{d} {N}}/{\text{d} E_{prompt}} = {\text{d} {N}}/{\text{d} E_{\vec{p}'}^f}\cdot({\text{d} E_{\vec{p}'}^f}/{\text{d} E_{prompt}}) $.
%The figure~\ref{Kam} shows predicted QES event induced by 0.1~MeV atmospheric DM in the KamLAND experiment, binned by deposited energy. The green and purple histograms correspond to atmospheric dark matter with incident energies of $E_{\chi} = 0.2$ and $0.6$~GeV, respectively. The details of the observed events in KamLAND are summarized in Table~\ref{table2} in the appendix, with contributions from well-predicted backgrounds and minor systematic effects already subtracted. 
The right plane of Fig.~\ref{Kam} shows predicted differential QES event induced by atmospheric DM with different mass in the KamLAND experiment. The orange, green and blue lines correspond to atmospheric dark matter with mass of $m_{\chi} = 1$, $10$, and $100$~GeV, respectively. As expected from Eq. \ref{eq:QES}, the event rates are suppressed at higer masses. Here, we have assumed perfect energy resolutions for the both experiments considered, while taking them into consideration has negligible effect \cite{Meighen-Berger:2023xpr}. 
%At high energies, effects such as PMT saturation limit the $E$ reconstruction ability of large liquid scintillators optimized for single-photon counting and $E$ resolution in the few-MeV range. At some point the energy reconstruction is expected to be highly nonlinear, and this will impact the ability of these experiments to extract the desired physics. 
{Following KamLAND \cite{KamLAND:2021gvi}, we apply the following selection criteria to the prompt and delayed events: prompt energy in the range 7.8--31.8~MeV, delayed energy in the range 1.8--2.6~MeV, spatial separation smaller than 160~cm, and time separation between 0.5 and 1000~$\mu$s. Owing to the one-to-one correspondence between prompt and delayed events, we present the observed events and the dominant expected background from atmospheric neutrino neutral-current interactions in KamLAND only as a function of the prompt energy, as shown in Table~1 in the Appendix. Contributions from well-understood backgrounds have already been subtracted~\cite{KamLAND:2021gvi,Meighen-Berger:2023xpr}.}
The systematic uncertainties associated with the neutrino fluxes and neutrino--nucleus cross sections are expected to be at the level of a few tens of percent~\cite{Super-Kamiokande:2015qek,Cheng:2020aaw,Cheng:2020oko}. We account for these effects by assigning a 25\% uncertainty to the background. 

% In the low deposited energy region of interest in this work, the dominant contribution comes from atmospheric DM with kinetic energies between 200 and 600 MeV. This is because higher-energy atmospheric DM is suppressed by kinematic constraints and spectral functions, while lower-energy DM will lack sufficient momentum transfer to effectively trigger the QES process.

\begin{figure}
\centering
\includegraphics[width=0.325\linewidth]{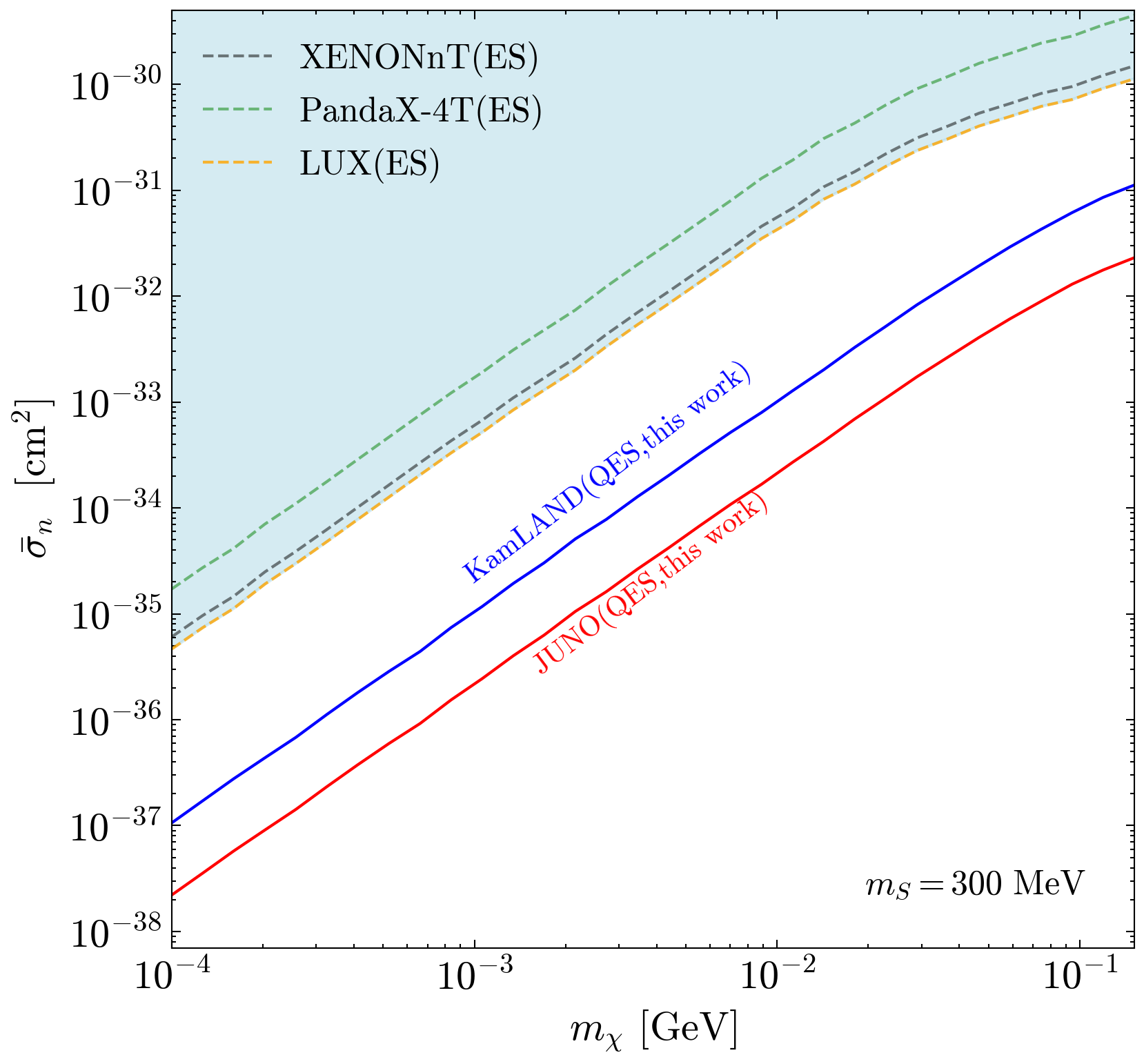}
\includegraphics[width=0.325\linewidth]{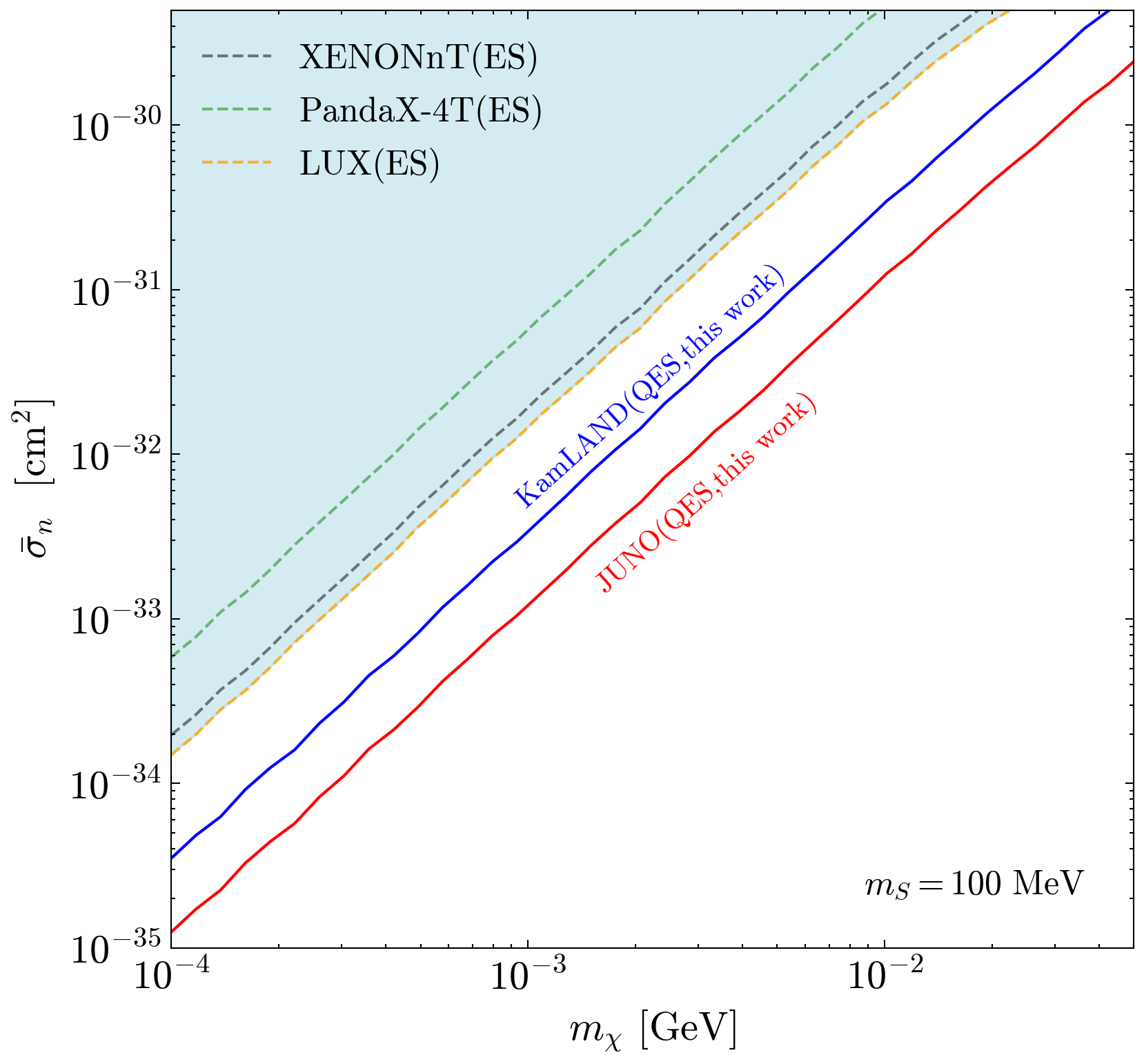}
\includegraphics[width=0.325\linewidth]{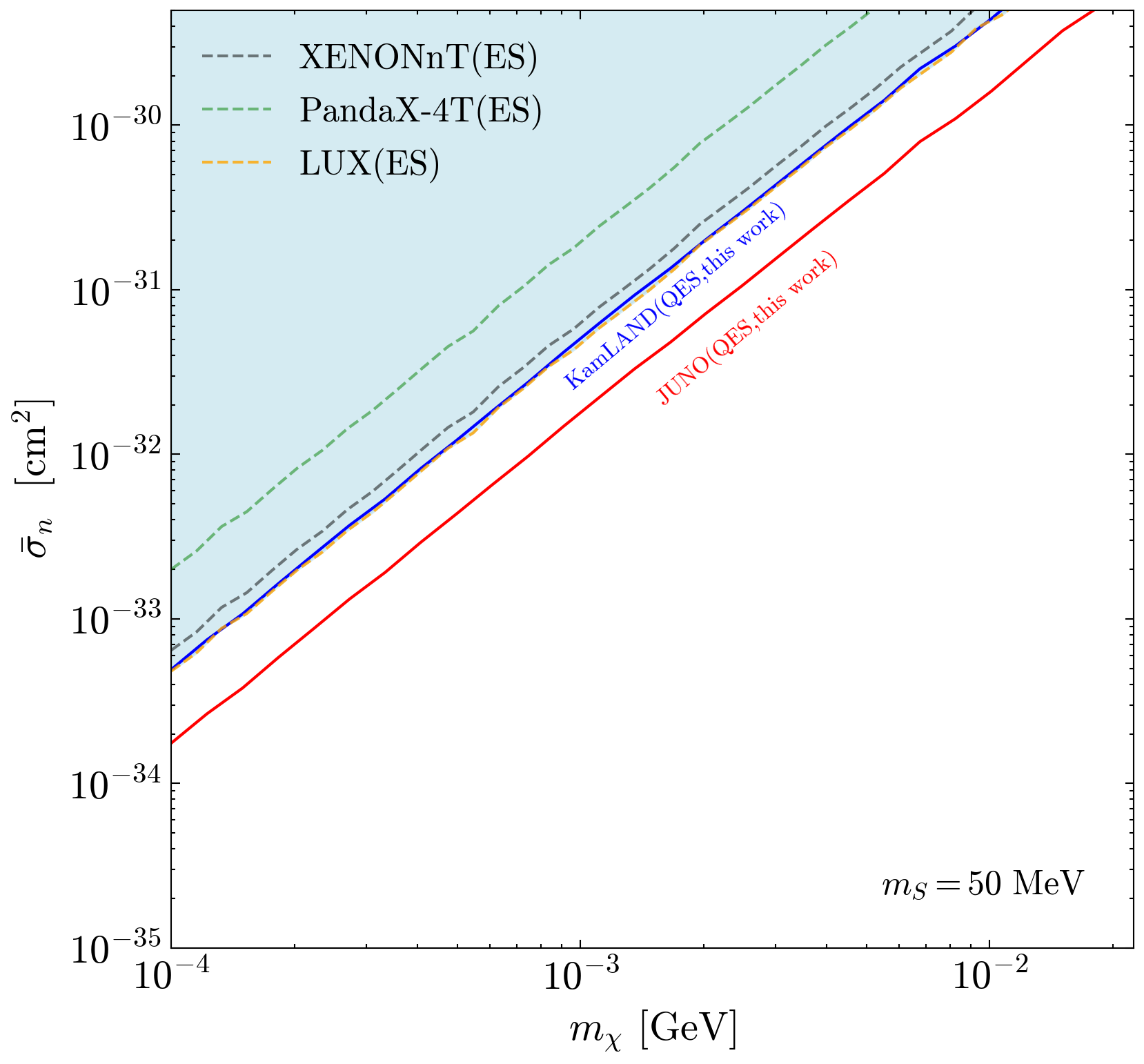}
\caption{90\% C.L. limits on the spin-independent atmospheric DM-nucleon scattering cross section versus the DM mass $m_\chi$ {for $m_S = 300$ MeV and BR$(\eta \to \pi \chi \bar{\chi})\ \simeq \ 1\times 10^{-5}$ (left plane), $m_S = 100$ MeV and BR$(\eta \to \pi \chi \bar{\chi})\ \simeq \ 6\times 10^{-7}$ (middle plane), and $m_S = 50$ MeV and BR$(\eta \to \pi \chi \bar{\chi})\ \simeq \ 6\times 10^{-7}$ (right plane), respectively}. The QES limits from KamLAND and a projection for the upcoming JUNO are shown in blue and the red lines, respectively. Other exclusion limits derived from ES processes are plotted by dashed lines [ LUX (orange), XENONnT (black), and PandaX-4T (green)].}
	\label{sigma_p}
\end{figure}

\section{Experimental sensitivity}

Utilizing the profile likelihood ratio approach, we derive the 90 \% C.L. exclusion limit on the atmospheric DM-nucleon  scattering cross section $\bar{\sigma}_n$ with the prompt-delayed events from KamLAND. We also present the expected 90\% C.L. sensitivity for the upcoming JUNO experiment at an exposure of 183 kt·yr, which bases on the predicted atmospheric neutrino events as background in the prompt energy range of 11–29 MeV~\cite{Meighen-Berger:2023xpr}. For comparison, we calculate the exclusion limits on $\bar{\sigma}_n$ with the data of elastic nuclear recoil from LUX, PandaX-4T and XENONnT provided in the appendix. The effective Lagrangian of 
DM-nucleus interactions can be described as $g_\chi S\bar{\chi}_L\chi_R+g_AS\bar{A}_LA_RF(Q^2)+\text{h.c.}$ ~\cite{Batell:2018fqo,AristizabalSierra:2018eqm}, where $g_A = Zg_{pS}+(A-Z)g_{nS}$ are the couplings of mediator $S$ with the nucleus $A$, $F(Q^2)$ is nuclear form factor ~\cite{Duda:2006uk}. Then, the coherent scattering cross section is given by~\cite{Su:2022wpj},
\begin{equation}
\begin{split}
\frac{\text{d}\sigma_{ES}}{\text{d}E_R}=&\frac{\bar{\sigma}_nA^2m_S^4F^2(E_R)}{32\mu_n^2m_A(2m_AE_R+m_S^2)^2(E_\chi^2-m_\chi^2)} (4m_\chi^2+2m_AE_R)(4m_A^2+2m_AE_R).
\end{split}
\end{equation}
The resulting constraints on $\bar{\sigma}_n$ for atmospheric DM are plotted in Fig.~\ref{sigma_p}. The detailed data and statistic method are provided in the Appendix.

We find that for $m_\chi = 0.1~(150)$ MeV, KamLAND has excluded the cross section above $1\times10^{-37}$ ($1\times10^{-31}$) cm$^2$ {for given $m_S = 300$ MeV and BR$(\eta \to \pi \chi \bar{\chi})\ \simeq \ 1\times 10^{-5}$ (left plane of Fig. \ref{sigma_p})}. This result is one order of magnitude more stringent than the those obtained from PandaX - 4T~\cite{PandaX:2024qfu}, XENONnT~\cite{XENON:2025vwd}, and LUX~\cite{LZ:2024zvo}. The projected sensitivity of JUNO demonstrates a five-fold improvement over KamLAND. 
It is noteworthy that the $m_S$ plays important roles in the QES and ES processes due to propagator effects, i.e., $\mathrm{d} \sigma \propto {1}/{\left(Q^2+m_S^2\right)^2}$. As discussed in Ref.~\cite{Su:2022wpj}, when the mediator is heavy, the QES cross section is larger than that of ES, making the QES process more favorable for detection. In the light mediator scenario, however, the $m_S$ term becomes negligible, resulting in an enhancement of the ES cross section proportional to $1/Q^4$. This enhancement narrows the gap between the exclusion limit of QES and ES process, {as can be seen from the plots for $m_S = 100$ MeV and BR$(\eta \to \pi \chi \bar{\chi})\ \simeq \ 6\times 10^{-7}$ (middle plane of Fig. \ref{sigma_p}), and $m_S = 50$ MeV and BR$(\eta \to \pi \chi \bar{\chi})\ \simeq \ 6\times 10^{-7}$ (right plane of Fig. \ref{sigma_p}). Compared to the heavier mediator scenario, the sensitivities for the two lighter mediator masses are weaker, primarily owing to the smaller branching fractions of the decay process $\eta \to \pi \chi \bar{\chi}$.}

\section{Conclusions}

The relativistic DM can reach Earth at sufficiently high velocities to induce the quasi-elastic scattering processes with nuclei. Large neutrino detectors are capable of identifying quasi-elastic signals that may arise from relativistic DM via prompt-delayed events. In this work, we consider the atmospheric DM scenario and calculate the differential cross section of atmospheric DM scattering off nucleus via a scalar mediator, $\chi + A \rightarrow \chi + (A-1)^{\star} + n$. With the prompt-delayed data from KamLAND experiment, we obtain the limits of the spin-independent DM-nucleon scattering cross section, which can be stronger than those from the elastic scattering processes measured in dark matter direct detection experiments. The upcoming JUNO experiment will further improve the sensitivity. In conclusion, the excellent energy resolution and neutron tagging capabilities of liquid scintillator detectors offer a significant advantage for observing the prompt-delayed events of knockout neutrons from the quasi-elastic scattering of DM-nucleus, which provides a new avenue for the detection of light dark matter.

%Finally, we can conclude that the prompt-delayed events from the quasi-elastic scattering of DM-nucleus provides a new avenue to detect the light DM in the liquid-scintillator neutrino experiments.

\section{Acknowledgments} 

This work is supported by the National Natural Science Foundation of China (NNSFC) under grant No. 12275134 and No. 12335005. The authors gratefully acknowledge the valuable discussions and insights provided by the members of the China Collaboration of Precision Testing and New Physics.

\section{Appendix}

\subsection{Data and Statistic Method} 

Tables \ref{table2} and \ref{table3} present the observed events versus expected backgrounds for QES and ES in different experimental configurations, respectively. The KamLAND data we use in this work is derived from~\cite{Meighen-Berger:2023xpr}, which, in comparison to the observed data in~\cite{KamLAND:2021gvi}, have had backgrounds from reactor, spallation, and atmospheric neutrino CC events subtracted, and the spectrum has been rebinned. The expected backgrounds arise primarily from atmospheric neutrino neutral-current interactions, for which a 25\% uncertainty is assigned to account for systematic uncertainties in the neutrino fluxes and neutrino--nucleus cross sections.
The profile likelihood ratio method was employed to analyze the data in this work. For KamLAND, we divide the observed events into 8 bins based on prompt energy and construct a histogram $\textbf{n}$. 
Assuming the event in each bin follow a Poisson distribution, the expected event count in the i-th bin can be expressed as $\textbf{\text{E}}[n_i]=s_i(\sigma)+b_i$, where $s_i(\sigma)$ corresponds to the signal hypothesis (with cross section $\sigma$) and $b_i$ represents known background contributions (atmospheric neutrino NC events here), including atmospheric neutrino interactions. The likelihood function, under the assumption is given by the product of Poisson probabilities for each bin,
\begin{equation}
{ L}_\sigma = \prod_i \frac{(s_i(\sigma) +b_i )^{n_i}}{n_i!}\mathrm{exp}[-(s_i(\sigma) +b_i )].
\end{equation}
We follow the method outlined in Ref ~\cite{Cowan:2010js} and utilize the test statistic $\tilde{q}_\mu$ and $q_0$ to establish the 90\% C.L. limit for KamLAND and JUNO experiment, respectively. Moreover, we use the pyhf package \cite{pyhf,pyhf_joss} to perfrom the likelihood test and incorporate the relevant uncertainties into the likelihood function.

\begin{table}[ht]
\begin{center}
\caption{Observed QES events and predicted backgrounds in KamLAND and JUNO ~\cite{Meighen-Berger:2023xpr}.}
\label{datasettable}
\begin{tabular}{ccccc}
\toprule
Exp &\ Prompt Energy (MeV)  & \ & Observed\ & \ Expt.bkg  \\
\midrule
KamLAND& $7.8- 10.8$  &\ & $4 \pm 2$ &3.9\\
(6.72 kt·yr)\ & $10.8- 13.8$  &\ & $2 \pm 1.5$ &3.1 \\
\ & $13.8- 16.8$  &\ & $2 \pm 1.5$ &2.9 \\
\ & $16.8- 19.8$  &\ & $2 \pm 1.5$ &2.2 \\
\ & $19.8- 22.8$  &\ & $2 \pm 1.5$ &2.1\\
\ & $22.8- 25.8$  &\ & $1 \pm 1$ &2\\
\ & $25.8- 28.8$  &\ & $1 \pm 1$ &1.9\\
\ & $28.8- 31.8$  &\ & $1 \pm 1$ &1.9\\
\midrule
JUNO & $11- 29$  &\ & $\setminus$ &412 \\
(183 kt·yr)\ & $\ $  &\ & $\ $ \\
\bottomrule
\label{table2}
\end{tabular}
\end{center}
\end{table}
\begin{table}[ht]
\begin{center}
\caption{Observed ES events and predicted backgrounds in different DM experiments~\cite{XENON:2025vwd,LZ:2024zvo,PandaX:2024qfu}.}
\label{datasettable}
\begin{tabular}{ccccc}
\toprule
Exp &\ Nuclear recoil  & \ & Observed\ & \ Expt.bkg  \\
\ & Energy (keV)   &\ & $\ $ \\
\midrule
XENONnT & $3.8- 64.1$  &\ & $397$ &$391 \pm 27$  \\
(2.4 t·yr)\ & $\ $  &\ & $\ $ \\
\midrule
LUX & $3.5- 65$  &\ & $1232$ &$1203 \pm 42$  \\
(3.3 t·yr)\ & $\ $  &\ & $\ $ \\
\midrule
PandaX-4T & $3- 103$  &\ & $2490$ &$2439 \pm 45$  \\
(1.54 t·yr)\ & $\ $  &\ & $\ $ \\
\bottomrule
\label{table3}
\end{tabular}
\end{center}
\end{table}

\bibliographystyle{JHEP}
\bibliography{refs.bib}


% \begin{thebibliography}{99}

% \bibitem{a}
% Author, \emph{Title}, \emph{J. Abbrev.} {\bf vol} (year) pg.

% \bibitem{b}
% Author, \emph{Title},
% arxiv:1234.5678.

% \bibitem{c}
% Author, \emph{Title},
% Publisher (year).

% Please avoid comments such as "For a review'', "For some examples",
% "and references therein" or move them in the text. In general,
% please leave only references in the bibliography and move all
% accessory text in footnotes.

% Also, please have only one work for each \bibitem.

% \end{thebibliography}
\end{document}